\documentclass[runningheads, envcountsame, a4paper]{llncs} 
\usepackage{inputenc}
\usepackage[centertags]{amsmath}
\usepackage{fullpage}
\usepackage{amstext,amssymb}
\usepackage{gastex}
\usepackage{todonotes}

\usepackage{xspace}
\newcommand{\bl}{\ensuremath{\left(}}
\newcommand{\br}{\ensuremath{\right)}}
\newcommand{\move}[1]{\ensuremath{\stackrel{#1}{\rightarrow}}}
\newcommand{\cmove}{\ensuremath{\rightarrow_c}}
\newcommand{\emove}{\ensuremath{\rightarrow_e}}
\newcommand{\rmove}{\ensuremath{\rightarrow_r}}
\newcommand{\moves}{\ensuremath{\twoheadrightarrow}}
\newcommand{\dmoves}[1]{\ensuremath{\stackrel{#1}{\moves}}}

\newcommand{\exptime}{{\sf EXPTIME}}

\newcommand{\tower}{{\sf Tow}}
\newcommand{\maxv}[1]{{\sf \ensuremath{MaxC_n(#1)}}}

\newcommand{\Ord}{{\ensuremath{\mathcal O}}}

\newcommand{\Equal}[1]{\ensuremath{{#1}-}{\sf Eq}}
\newcommand{\Valid}[1]{\ensuremath{{#1}-}{\sf Val}}
\newcommand{\Succ}[1]{\ensuremath{{#1}-}{\sf Succ}}
\newcommand{\last}[1]{\ensuremath{{\sf maxval(#1)}}}
\newcommand{\first}[1]{\ensuremath{{\sf minval(#1)}}}
\newcommand{\error}{\ensuremath{q_{err}}}
\newcommand{\ValidConf}[1]{\ensuremath{{#1}-}{\sf ValConf}}
\newcommand{\Move}[1]{\ensuremath{{#1}-}{\sf ValMov}}
\newcommand{\tmoves}[1]{\ensuremath{\vdash_{#1}}}
\newcommand{\InitConf}[2]{\ensuremath{{(#1,#2)}-}{\sf InitConf}}
\newcommand{\FinalConf}[1]{\ensuremath{{#1}-}{\sf FinalConf}}
\newcommand{\EqConf}[1]{\ensuremath{{#1}-}{\sf EqConf}}
\newcommand{\SuccConf}[1]{\ensuremath{{#1}-}{\sf SuccConf}}

\newcommand{\EX}{\ensuremath{\mathrm{EX}}}
\newcommand{\EF}{\ensuremath{\mathrm{EF}}}
\newcommand{\EG}{\ensuremath{\mathrm{EG}}}
\newcommand{\EU}{\ensuremath{\mathrm{EU}}}

\newcommand{\qmax}[1]{\ensuremath{q^{max?}_{#1}}}
\newcommand{\qmin}[1]{\ensuremath{q^{min?}_{#1}}}
\newcommand{\ctlmax}[1]{\ensuremath{\Phi^{ctr}_{max}(#1)}}
\newcommand{\ctlmin}[1]{\ensuremath{\Phi^{ctr}_{min}(#1)}}
\newcommand{\qeqcheck}[1]{\ensuremath{q^=_{in,#1}}}
\newcommand{\Qeqcheck}[1]{\ensuremath{Q^=_{#1}}}
\newcommand{\Qskip}[1]{\ensuremath{Q^{skip}_{#1}}}
\newcommand{\qskipfinal}[1]{\ensuremath{q^{skip}_{f,#1}}}
\newcommand{\qrcchk}[1]{\ensuremath{q^{rc,=}_{#1}}}
\newcommand{\Qrcchk}[1]{\ensuremath{Q^{rc,=}_{#1}}}
\newcommand{\ctleq}[1]{\ensuremath{\Phi^{ctr}_{=}(#1)}}
\newcommand{\qsuccheck}[1]{\ensuremath{q^{+1}_{in,#1}}}
\newcommand{\Qsuccheck}[1]{\ensuremath{Q^{+1}_{#1}}}
\newcommand{\ctlsucc}[1]{\ensuremath{\Phi^{ctr}_{+1}(#1)}}
\newcommand{\qskipfinalsuc}[1]{\ensuremath{q^{skip+}_{f,#1}}}
\newcommand{\qrcchkneq}[1]{\ensuremath{q^{rc,\neq}_{#1}}}
\newcommand{\qscantype}[1]{\ensuremath{q^{type}_{#1}}}
\newcommand{\qeq}{\ensuremath{q_{eq}}}
\newcommand{\qneq}{\ensuremath{q_{neq}}}
\newcommand{\qvalid}[1]{\ensuremath{q^{val}_{in,#1}}}
\newcommand{\Qvalid}[1]{\ensuremath{Q^{val}_{#1}}}
\newcommand{\ctlval}[1]{\ensuremath{\Phi^{ctr}_{val}(#1)}}
\newcommand{\qvalskip}[1]{\ensuremath{q^{1-skip}_{in,#1}}}
\newcommand{\qvalinduct}[1]{\ensuremath{q^{C}_{#1}}}
\newcommand{\ctlvalind}[1]{\ensuremath{\Phi^{ctr}_{val,ind}(#1)}}
\newcommand{\qvalskipp}[1]{\ensuremath{p^{1-skip}_{in,#1}}}
\newcommand{\qvalinductp}[1]{\ensuremath{p^{C}_{#1}}}
\newcommand{\qchklast}[1]{\ensuremath{q^{last}_{#1}}}
\newcommand{\win}{\ensuremath{q_{win}}}
\newcommand{\ctlvallast}[1]{\ensuremath{\Phi^{ctr}_{val,last}(#1)}}
\newcommand{\ctlvalfirst}[1]{\ensuremath{\Phi^{ctr}_{val,first}(#1)}}
\newcommand{\qvalskipr}[1]{\ensuremath{r^{1-skip}_{in,#1}}}
\newcommand{\qguess}[1]{\ensuremath{q^{guess}_{#1}}}
\newcommand{\qchkguess}[1]{\ensuremath{q^{guess}_{#1,chk}}}
\newcommand{\qremovelj}[1]{\ensuremath{q^{rm}_{l_j,#1}}}
\newcommand{\qbeforesucccheck}[1]{\ensuremath{q^{+1'}_{in,#1}}}
\newcommand{\ctlvalsucc}[1]{\ensuremath{\Phi^{ctr}_{val,succ}(#1)}}

\newcommand{\qvalidconf}[1]{\ensuremath{q^{valC}_{in,#1}}}
\newcommand{\ctlvalidconf}[1]{\ensuremath{\Phi^{con}_{val}(#1)}}
\newcommand{\qinit}[2]{\ensuremath{\ensuremath{q^{initC}_{#1,#2}}}}
\newcommand{\ctlinitconf}[2]{\ensuremath{\Phi^{con}_{init}(#1,#2)}}
\newcommand{\qfinal}[1]{\ensuremath{\ensuremath{q^{finalC}_{#1}}}}
\newcommand{\ctlfinalconf}[1]{\ensuremath{\Phi^{con}_{final}(#1)}}
\newcommand{\qeqconfig}[1]{\ensuremath{q^{con,=}_{in,#1}}}
\newcommand{\ctleqconf}[1]{\ensuremath{\Phi^{con}_{=}(#1)}}

\newcommand{\qsimple}[1]{\ensuremath{q^{nm,s}_{in,#1}}}
\newcommand{\qsimplef}[1]{\ensuremath{q^{nm,s}_{f,#1}}}
\newcommand{\qnotfromq}[1]{\ensuremath{q^{\neg XQ}_{in}}}
\newcommand{\rsimple}[1]{\ensuremath{r^{nm,s}_{in,#1}}}
\newcommand{\rsimplef}[1]{\ensuremath{r^{nm,s}_{f,#1}}}
\newcommand{\qrcchkn}[1]{\ensuremath{q^{rc}_{#1}}}
\newcommand{\ctlsimple}[1]{\ensuremath{\Phi^{con}_{\vdash,s}(#1)}}

\newcommand{\qhard}[1]{\ensuremath{q^{nm,h}_{in,#1}}}

\newcommand{\qtrymoves}[1]{\ensuremath{q^{mov}_{#1}}}
\newcommand{\rhard}[1]{\ensuremath{r^{nm,h}_{in,#1}}}
\newcommand{\rhardf}[1]{\ensuremath{r^{nm,h}_{f,#1}}}
\newcommand{\rtrymoves}[1]{\ensuremath{r^{mov}_{#1}}}
\newcommand{\ctlhard}[1]{\ensuremath{\Phi^{con}_{\vdash,h}(#1)}}
\newcommand{\qmove}[1]{\ensuremath{q^{nm}_{in,#1}}}
\newcommand{\ctlmove}[1]{\ensuremath{\Phi^{con}_{\vdash}(#1)}}

\newcommand{\qguessconfchk}[1]{\ensuremath{q^{g,con}_{#1,chk}}}
\newcommand{\qrcmove}[1]{\ensuremath{q^{rcon}_{#1}}}
\newcommand{\qrcmovef}[1]{\ensuremath{q^{rcon}_{#1,f}}}
\newcommand{\qstep}[1]{\ensuremath{q^{\vdash}_{in,#1}}}
\newcommand{\ctlstep}[1]{\ensuremath{\Phi_{\vdash}(#1)}}

\newcommand{\qinitcheck}[2]{\ensuremath{q^{initC}_{chk}(#1,#2)}}
\newcommand{\qmovecheck}[1]{\ensuremath{q^{move}_{chk}(#1)}}
\newcommand{\qwrite}[1]{\ensuremath{q_{guessC}(#1)}}
\newcommand{\qstart}[1]{\ensuremath{q_{acc}(#1)}}
\newcommand{\ctlaccept}[2]{\ensuremath{\Phi^M_{acc}(#1,#2)}}

\newcommand{\qruncheck}[1]{\ensuremath{q^{run}_{#1}}}
\newcommand{\qremoveconf}[1]{\ensuremath{q^{remC}_{#1}}}
\newcommand{\qonemore}[1]{\ensuremath{q^{1+}_{#1}}}
\newcommand{\qtwomore}[1]{\ensuremath{q^{2+}_{#1}}}
\newcommand{\ctlacceptp}[2]{\ensuremath{\Psi^M_{acc}(#1,#2)}}
\newcommand{\Comment}[1]{}

\newcommand{\FO}{\ensuremath{{\rm FO}(<)}}
\newcommand{\Nat}{\ensuremath{\mathbb{N}}}
\newcommand{\Qx} {\ensuremath{\mathcal Q}x}
\newcommand{\Subf}[1]{\ensuremath{\mathcal S(#1)}}

\newcommand{\True}{{\sf T}}
\newcommand{\False}{{\sf F}}

\title{Model checking Branching-Time Properties of Multi-Pushdown Systems is Hard}

\author{Mohamed Faouzi Atig\inst{1} \and Ahmed  Bouajjani\inst{2} \and K. Narayan Kumar\inst{3} \and Prakash Saivasan\inst{3}}
\authorrunning{M. F. Atig, A. Bouajjani, K. Narayan Kumar and P. Saivasan}

\institute{
    Uppsala University, Sweden / \email{mohamed\_faouzi.atig@it.uu.se}
    \and
    LIAFA, Universit\'e Paris Diderot, France /
    \email{abou@liafa.univ-paris-diderot.fr}
    \and
    Chennai Mathematical Institute, India / \email{\{kumar,saivasan\}@cmi.ac.in}
}

\begin{document}

\maketitle

\begin{abstract}
We address the model checking problem  for shared memory concurrent programs modeled as multi-pushdown systems. We consider here boolean programs with a finite number of threads and recursive procedures. It is well-known that the model checking problem is undecidable for this class of programs. In this paper, we investigate the decidability and the complexity of this problem under the assumption of  bounded context-switching defined by Qadeer and Rehof \cite{QaRe05},  and  of phase-boundedness proposed  by  La Torre et al.  \cite{ToMaPa07}.  On the model checking of such systems against temporal logics and in particular
branching time logics such as the modal $\mu$-calculus or CTL has received little
attention.
It is known that parity games, which are closely related to the modal $\mu$-calculus, 
are decidable for the class of bounded-phase systems (and hence for bounded-context switching as well), but with non-elementary complexity \cite{Se09}.  
A natural question is whether this high complexity is inevitable and what
are the ways to get around it. This paper addresses these questions and
unfortunately, and somewhat surprisingly,  it shows  that branching model 
checking for MPDSs is inherently an hard problem with no easy solution. We show that  parity games on MPDS under phase-bounding restriction  is  non-elementary. Our main result shows that model checking a $k$ context bounded MPDS 
against a simple fragment of CTL, consisting of formulas that whose 
temporal operators come from the set $\{\EF, \EX\}$,  has a non-elementary 
lower bound.

\end{abstract}

\section{Introduction}

The verification of multi-threaded programs is an important topic
of research in recent years \cite{AtTo09,AtBoQa09,HeLeMuSu10,Ka09,LaRe09,LaToKiRe08,QaRe05,ToMaPa08a}. One may use pushdown systems to abstract
sequential recursive programs and analyze them using the plethora
of results available in literature. However, the presence of multiple-threads
with their own call stacks means that modeling multi-threaded programs
needs systems with multiple pushdowns. Unfortunately, verifying 
a finite state system equipped in addition with 2 pushdowns is undecidable
as it is turing powerful. 

Qadeer and Rehof \cite{QaRe05} proposed one way to get around this undecidability.
They studied under-approximations of the set of behaviors of 
multi-pushdown systems. They proposed the \emph{bounded context-switching}
restriction, that imposes a bound $k$ on the number of times of  switches 
from using one pushdown to another. The control state reachability 
as well as the global model checking problem (computing, for a given
regular set of configurations, the set of configurations from which the
given set can be reached)  turn out to be decidable. 
Subsequently, various other classes of under-approximations have been
studied including bounded phase, ordered multi-pushdown and bounded scope
\cite{At10b,At10a,AtBoHa08,BrChCiCr96,MaPa11,Se10,ToMaPa07,ToNa11}.

A phase is a sequence of computational steps that pops from a fixed stack
but is allowed to push values into any stack. By imposing a bound $k$ on the
number of phases, we obtain an under-approximation that is more general than
 bounded context switch analysis. This restriction, called the bounded-phase restriction
was proposed in \cite{ToMaPa07}, where its controls state reachability
problem is also shown to be decidable. 

In \cite{BrChCiCr96,AtBoHa08} a different restriction called ordered multi-pushdown is studied
where there is  linear order on the stack and any pop action is only
permitted in the smallest nonempty stack. More recently, in \cite{ToNa11}, 
a restriction that demands that a value that is pushed be popped within a
bounded number of context switches (or not at all) is studied.
Most of these works examine the control state reachability problem
and its generalization, the global reachability problem and obtain
decidability results \cite{At10b,Se10}.

On the model checking of such systems against temporal logics and in particular
branching time logics such as the modal $\mu$-calculus or CTL has received little
attention.
It is known that parity games, which are closely related to the modal $\mu$-calculus, 
are decidable for the class of bounded-phase systems (and hence for bounded-context switching as well), but with non-elementary complexity \cite{Se09}.  
A natural question is whether this high complexity is inevitable and what
are the ways to get around it. This paper addresses these questions and
unfortunately, and somewhat surprisingly,  it seems that branching model 
checking for MPDSs is inherently an hard problem with no easy solution. 
Our main result shows that model checking a $k$ context bounded MPDS 
against a simple fragment of CTL, consisting of formulas that whose 
temporal operators come from the set $\{\EF, \EX\}$,  has a non-elementary 
lower bound.

The complexity of parity games and CTL model-checking for pushdown systems
has been well studied. Walukiewicz \cite{Wa01} shows that parity games
are solvable in EXPTIME and that model checking of PDSs against even
CTL formulas has a EXPTIME lower-bound \cite{Wa00}. As a matter of fact, our 
proof utilizes ideas from the latter work.

A different generalization of pushdown systems is that of higher-order pushdown
systems (HOPDAs). A level 1 pushdown is a normal pushdown and a level  $k$ pushdown
has a pushdown of level $k-1$ pushdowns. A higher level push operation duplicates
 the top most stack while a pop operation removes such a stack.  For a formal definition of these
models and the operations on them the reader is referred to \cite{Wo05,CaWo03}.  
These are extremely powerful models and in \cite{CaWo03} it is shown that 
their configuration graphs capture every graph that lies in the \emph{Caucal 
hierarchy}. Cachat \cite{Ca03a} also showed the decidability of parity games
over HOPDAs.  Cachat and Walukiewicz \cite{CaWa07} show that parity games on 
HOPDAs has non-elementary complexity on the number of levels of higher order stacks and subsequently tight lower bounds have been shown for the model checking of HOPDAs w.r.t. various linear and branching time temporal logics \cite{HaTo10}.
 A key ingredient in the lower bound proof of Cachat-Walukiewicz  is the 
use of a certain kind of counters, introduced by L. Stockmeyer \cite{St74}, 
and encoding of the configurations of a TM using these counters. We draw
heavily on this idea in our lower bound proof for CTL. Unlike the HOPDAs,
bounded context switch MPDSs do not posses the ability to duplicate the
contents of a stack making our argument somewhat more elaborate. 

\section{Preliminaries}

A multi-pushdown system (MPDS) is a generalization of the classical pushdown system
with multiple stacks. As it is well known, two stacks suffice  to simulate a 
tape and hence even a two stack MPDS is turing powerful.  However, there
are a number of restrictions that one may place  the behaviors of MPDSs resulting in decidability of many interesting
properties.

\begin{definition}
\label{defn:MPDS}
A Multi Pushdown System MPDS  $A$ is a tuple $\bl Q, \Gamma, l, \delta, q_0
\br$ where $Q$ is a finite set of states, $l$ is an integer giving the number 
of stacks, $\Gamma$ is the stack alphabet (not containing the special stack symbol $\bot)$, $q_0$ is the initial state and $\delta$ = $\delta_e \;
\cup \; \delta_c \; \cup \; \delta_r$ is the transition relation, where
\begin{itemize}
  \setlength{\itemsep}{1pt}
  \setlength{\parskip}{0pt}
  \setlength{\parsep}{0pt}
\item[-] $\delta_e \subseteq Q \times Q$
\item[-] $\delta_c \subseteq Q \times (\Gamma \cup \{\bot\}) \times Q  \times [ 1..l ] \times \Gamma$ 
\item[-] $\delta_r \subseteq Q \times \Gamma \times Q \times [1 .. l ] $
\end{itemize}
\end {definition}

In each transition, the MPDS may carry out an internal (or \emph{skip}) move 
($\delta_e$), or examine the top symbol of one stack and 
based on its value a \emph{push} one symbol that stack ($\delta_c$)
or a \emph{pop} one symbol from that stack ($\delta_r$).
We shall write $\delta_r^i$, $1 \leq i \leq l$, to denote the set of
pop transitions where the pop is performed on stack $i$ and similarly
$\delta_c^i$ will denote the set of push transitions on stack $i$.
The configuration of such a MPDS  is naturally given by the current state
as well as the contents of the $l$ stacks.
\begin{definition}
\label{defn:configuration}
A \emph{configuration} of a MPDS $A = \bl Q, \Gamma, l, \delta, q_0 \br$  is of the form  $q\bl \gamma_1, \cdots, \gamma_l \br$ where $q \in Q$ is a state and $\gamma_i \in  \Gamma^* \cdot \{\bot\}$ is the content of the stack $i \in [1..n]$. 
\end{definition}

Next we define the one step move relation which describes how an MPDS
may move from one configuration to another using one of the transitions
in $\delta$.

\begin{definition}
\label{defn:move}
Let $A = \bl Q, \Gamma, l, \delta, q_0 \br$ be a MPDS.
The one step move relation  using the transition $t \in \delta$ is defined
as follows:

$$q\bl \gamma_1 , \cdots ,\gamma_l \br \quad \stackrel{t}{\rightarrow} \quad q'\bl \gamma ' _1 , \cdots, \gamma' _l \br$$

if and only if one of the following conditions holds

\begin{enumerate}
  \setlength{\itemsep}{1pt}
  \setlength{\parskip}{0pt}
  \setlength{\parsep}{0pt}
     \item { $t=\bl q,q' \br\in \delta_e$ and $\gamma_{i} = \gamma'_i$.}
     \item { $t=\bl q,a,q',j,b \br$ $\in$ $\delta_c$ and  $\gamma'_j = b.\gamma_{j}$, 
$\gamma_j = a.\gamma$  and for $i \neq j$,  $\gamma_{i} = \gamma'_i$ }
     \item { $t=\bl q,a,q',j \br$ $\in$ $\delta_r$ and  $\gamma'_j = \gamma_{j} $ and $\gamma_j = a.\gamma_j$ and for $i \neq j$,  $\gamma_{i} = \gamma'_i$ }
\end{enumerate}
\end{definition}

\paragraph{Notation}
We write $\move{}$ to denote $\bigcup_{t \in \delta} \move{t}$, 
$\emove$ to denote $\bigcup_{t \in \delta_e} \move{t}$, $\cmove$ to denote $\bigcup_{t \in \delta_c} \move{t}$ and $\rmove$ to denote $\bigcup_{t \in \delta_r} \move{t}$.
We use the $\moves$  to denote the reflexive, transitive closure
of $\rightarrow$. We also  write $\dmoves{w}$ with
$w \in \delta^*$, when the sequence of transitions used is important.
We say that there is a \emph{run} from a configuration $c$ to a configuration
$d$ if $c \moves d$ and that there is a run over $w$ ($w \in \delta^*$) if
$c \dmoves{w} d$.

We write $\moves_i$ to denote the reflexive transitive
closure of $\delta_e \cup \delta_c^i \cup \delta_r^i$, i.e. sequences
of moves in which all stack accesses are restricted to the stack $i$. 
We also use $\delta^i$ to denote the set $\delta_e \cup \delta_c^i \cup \delta_r^i$.

Informally, a \emph{context} is a sequence of moves in which only a single stack
is accessed. Clearly, each run of an MPDS 
can be broken up into contiguous segments, where each segment forms a context.
Qadeer and Rehof \cite{QaRe05} in 2005, showed that by a priori bounding the 
number of contexts in any execution by a constant $k$ (or equivalently by restricting our
attention only to runs whose number of contexts is bounded by a constant $k$) one
can effectively analyze multi-pushdown systems. For instance, the control state reachability
problem becomes decidable.

\begin{definition}
\label{defn:mcontextreach}
Let $c$ be a configuration. A run $c \dmoves{w} d$ is said
to be $m$-context if $w = w_1. w_2. w_3 \ldots w_m$ such that
for each $j$ with $1 \leq j \leq m$, there is an $i_j$, $1 \leq
i_j \leq l$, such that $w_j \in ({\delta^{i_j}})^*$.
We say that $d$ is reachable from $c$ in $m$ context switches if there
is a $w$ and a $m$-context run $c \dmoves{w} d$.
\end{definition}

The idea of a context can be generalized to a \emph{phase} by focussing
only on the pop moves in the run. In a \emph{phase} of a run of an MPDS, 
all the  pop moves involve the same stack.  Each run of an MPDS can be broken up into contiguous segments, 
wherein each segment forms a phase. The bounded-phase restriction places
a bound $k$ on the number of phases along any run.

\begin{definition}
\label{defn:mphasereach}
Let $c$ be a configuration. A run $c \dmoves{w} d$ is said
to be $m$-phase, if $w = w_1. w_2. w_3 \ldots w_m$ such that 
for each $j$ with $1 \leq j \leq m$, there is an $i_j$, $1 \leq
i_j \leq l$, such that $w_j \in (\delta^{i_j}\cup \bigcup_{p \leq l} \delta^p_c)^*$.
Finally, $d$ is reachable from $c$  in $m$ phases if there
is a $w$ such that $c \dmoves{w} d$ is $m$-phase.
\end{definition}

\section{Parity Games over MPDSs}

We now define parity games over MPDSs and subsequently consider their
restriction to bounded number of phases.

\begin{definition}
\label{defn:ParityMPGame}
A parity game over an MPDS is a MPDS $A = (Q, \Gamma, l, q_0, \delta)$, along
with a decomposition $Q$ into two disjoint sets $Q_0$ and $Q_1$ (i.e., $Q = Q_0 \uplus Q_1$)  and a 
a ranking function $\Omega : Q \longrightarrow [1..M]$.  
The positions
of such a game are the configurations of the MPDS. A position
$q(\gamma_1,\gamma_2, \cdots, \gamma_l)$ belongs to player $i$ if $q$ belongs to $Q_i$
and its rank is $\Omega(q)$.   
Since the starting state of the MPDS  often plays no role in the 
definition of games we shall usually drop it from the definition of 
MPDS in the following and write a game $G$ as a pair $(A,\Omega)$ where
$A = (Q_0 \uplus Q_1, \Gamma, l, \delta)$ is an MPDS (w/o a start state) and
$\Omega$ is a ranking function.

The usual notions of plays, strategies, winning strategies,  memoryless strategies, plays consistent
with a given strategy and so on are defined on these game
graphs as they are just a subclass of parity games.
\end{definition}

Classical theorems such as Martin's determinacy theorem as well 
as the memoryless determinacy theorem hold for these games  as the
winning condition is a parity condition. However, since MPDSs with even
two stacks are Turing powerful it follows that there is no hope for
algorithmic solvability. 

In \cite{Se09} Anil Seth showed that parity games on MPDSs with a bound on the
number of phases is decidable.

\begin{definition}
\label{defn:kPhaseGame}
Let $G = (A,\Omega)$ be a MPDS parity game where $A=(Q_0 \uplus Q_1,\Gamma,l,\delta)$.
 The positions of the bounded-phase game on $G$ are triples of the form $(c,i,k)$ where $c$ is a configuration of the MPDS $A$, $i \in \{0,1,\cdots,l\}$ is
 a stack identifier and $k > 0$ is an integer denoting the remaining number
 of phases. The number $k$ indicates an upper bound on the number of phases 
 that are permitted starting at the configuration $c$ and the number $i$ gives
 the stack being used in the current phase. The value $i=0$ is used to 
 indicate that the current phase has not used any stack (this is
the case at the beginning of the game). The edges of the game graph
 are given by $(c,i,k) \move{} (c',i',k')$ if 
 \begin{enumerate}
  \setlength{\itemsep}{1pt}
  \setlength{\parskip}{0pt}
  \setlength{\parsep}{0pt}
  \item  $c \emove c'$ or $c \cmove c'$ and $i' = i$ and $k' = k$
  \item  $c \move{t} c'$, $t \in \delta^j_r$,  $i=0$, $k=k'$ and $i' = j$
  \item  $c \move{t} c'$, $t \in \delta^i_r$, $k=k'$, $i'=i$
  \item  $c \move{t} c'$, $t \in \delta^j_r$, $j \neq i$, $k > 1$, $k' = k-1$, $i'=j$.
 \end{enumerate}

 Observe that if the game is already in a position of the form $(c,i,1)$ 
 then pop moves on any stack other than $i$ are no longer available. Thus, even
 if the original MPDS has no deadlocked configurations, the game graph
 described above might still have positions with no outgoing edges. 
 As usual, if the game reaches a position with no outgoing edges then
 the owner of that position loses the game.

 The ranking function assigns ranks based on the local state of the MPDS
 $$\Omega(q(\gamma_1, \gamma_2, \cdots,\gamma_l),i,k) = \Omega(q)$$
 \end{definition}

 We say that a player $i$ wins the $k$-phase game starting at a configuration
 $c$ of the MPDS $A$, if the position $(c,0,k)$ is winning in the game 
 graph described above.  Anil Seth proved the following theorem:

\begin{theorem}(Anil Seth) The MPDS  parity game  with a phase bound $k$   is  decidable. That is, one can determine for any starting configuration
$c$ the winner from that position. 
\end{theorem}

The construction in \cite{Se09} also shows that the winner's strategy can 
be described as a multi-pushdown strategy. The complexity of determining
the winner is non-elementary and grows as a tower of exponentials as $k$
increases. As our first result, in the next section, we show that this
is inevitable by establishing a non-elemenatry lower bound for such games,
there by settling an open question posed in \cite{Se09}.

A natural question then is consider weaker models (than bounded phase systems)
or weaker properties (than parity games, which are equivalent to the modal
$\mu$-calculus) or both. Surprisingly, we find that even for the weakest
model of MPDSs considered, with a bound $k$ on the number of context switches,
and a fragment of the logic CTL, which in turn is a simple fragment of 
the modal $\mu$-calculus, the model checking problem turns out to be
non-elementary and grows as a tower whose height grows linearly in $k$.
This proof is significantly more complicated and draws heavily from the
techniques developed in \cite{Wa00} by Walukiewicz and in \cite{CaWa07} by
Cachat and Walukiewicz. The rest of the paper describes
a proof of this result.

\section{A lower bound for bounded-phase parity games}

A well known result of Stockmeyer \cite{St74} shows that deciding the
satisfiability of the  first order logic with the ordering relation ($\FO$) 
over $(\Nat, <)$ (or the validity, since validity is the same as
satisfiability over a single model) has non-elementary complexity.

We now show that given a formula $\phi$ in $\FO$ of size $n$ and 
quantifier  depth $k$ (clearly $k \leq n$) there is an MPDS 
that is polynomial in size of $\phi$ such that the $k$ phase game is 
winning for player $0$ if and only if the formula $\phi$ is satisfiable.

Henceforth we assume that there are no negations in the formula
(this can be ensured by pushing the negations down to the atomic
formulas using the usual dualties and then replacing $\neg (x < y)$ by
$(x = y) \lor (y < x)$ and so on.

\subsection{The satisfiability game}

We define a reachability game whose positions are pairs of the form 
$(\psi,\rho)$, 
where $\psi$ is a formula from $(\FO)$ and $\rho:FV(\psi) \rightarrow \Nat$
is a function that assigns a  natural number  to each of the free variables of
$\psi$. If the outer most logical operator of $\psi$ is either a $\forall$
quantifier or $\land$ then the  position of the form $(\psi,\rho)$ belongs
to player 1. Otherwise, i.e. if the outermost logical operator is either a
$\exists$ quantifier or $\lor$ or the formula is an atomic formula then
the position belongs to player $0$. 

If $\psi$ is an atomic formula then it has no outgoing edges. If $\psi$
is $\psi_1 \lor \psi_2$ or $\psi_1 \land \psi_2$ then there are edges
from any position of the form $(\psi,\rho)$ to the positions $(\psi_1,\rho)$
and $(\psi_2,\rho)$. If $\psi = \forall x. \psi'$ (or $\psi = \exists x. \psi')$
then there are edges from $(\psi,\rho)$ to all positions of the form
$(\psi',\rho')$ where $\rho'(y) = \rho(y)$ for $y \neq x$ and $\rho'$ is
also defined at $x$.

The play is winning for player $0$ if it ends at a node  
of the form $((x = y), \rho)$ and $\rho(x) = \rho(y)$ or it ends at a node
of the form $((x < y), \rho)$ and $\rho(x) < \rho(y)$.  Otherwise, player
$1$ wins the game.  The following is quite easy to see.

A winning strategy for player $0$ picks positions for the 
existential variables in such a way that no matter which positions are 
picked for the universal variables by the opponent the resulting 
quantifier-free formula is satisfied.  It is easy to see that 

\begin{theorem}\label{thm:satgame}
Given a formula $\phi$  and a valuation $\rho$ for the free variables
of $\phi$, $\phi$ is satisfiable/valid  w.r.t. $\rho$  iff player 0
has a winning strategy from the position $(\phi,\rho)$ in the satisfiability
game. In particular, if $\phi$ is a sentence then 
it is satisfiable/valid iff player $0$ has a winning strategy from the
position $(\phi,\emptyset)$.
\end{theorem}

\subsection{The bounded phase game for $\FO$ satisfiability}
We now show that the satisfiability game can be reformulated as a 
bounded-phase MPDS game. Let $\phi$ be the given formula. Informally,
the MPDS maintains the current valuation $\rho$ in its first stack
and the formula $\phi$ in the state. In each step, the automaton strips
off one operator from the formula. Stripping a quantifier corresponds
to modifying the contents of the stack to reflect the new valuation. 

 We translate a valuation $\rho$ into a word as follows: If the domain of $\rho$ is empty then we 
represent it using the empty word. Otherwise, it is represented by
any word $w$ over the alphabet $\{a\} \cup V$ where $V$ is the domain
of $\rho$, $w \in V \cdot (\{a\} \cup V)^*\cdot  \{\bot\} $, every element of $V$ occurs
precisely once in $w$ and   if $w = w_1 x w_2$ for $x \in V$ then
$\#_a w_2 = \rho(x)$. 

Let $\Subf{\phi}$ be the set of sub-formulas of the formula $\phi$ and let
$V$ be its set of variables. We describe the MPDS in two parts. The first
part describes the moves till we reach an atomic formula.    The 
set of states of used for this purpose is 
$\Subf{\phi} \cup (\Subf{\phi} \times \{>, <,1t2,2t1\}) \cup (\Subf{\phi} \times \{1t2,2t1\} \times (\{a\} \cup V))$. 
The transitions are defined as follows (we write $\Qx$ to stand for $\forall x$
and $\exists x$):
\begin{enumerate}
  \setlength{\itemsep}{1pt}
  \setlength{\parskip}{0pt}
  \setlength{\parsep}{0pt}
\item $(\psi_1 \land \psi_2, \psi_1), (\psi_1 \land \psi_2, \psi_2) \in \delta_e$.
\item $(\psi_1 \lor \psi_2, \psi_1), (\psi_1 \lor \psi_2, \psi_2) \in \delta_e$.
\item $(\Qx.\psi, (\Qx.\psi,<)), (\Qx. \psi,(\Qx.\psi,>))  \in \delta_e$. Guess whether the
next variable $x$  is to be inserted between existing variables or to their right.
\item $((\Qx.\psi,>),., (\Qx.\psi,>),1,a) \in \delta_c$. Push an $a$ to increase the possible   number for $x$. (Observe that we use the symbol $.$ to denote that there is no constraint on the top of the stack.)
\item $((\Qx.\psi,>).,\psi,1,x)\in \delta_c$. Mark the position for $x$ and shift to the sub-formula.
\item $((\Qx.\psi,<),(\Qx.\psi,1t2) \in \delta_e$.  Begin copying some elements from Stack 1 to Stack 2.
\item $((\Qx.\psi,1t2),c, (\Qx.\psi,1t2,c), 1) \in \delta_r$. Read and pop a value from stack 1.
\item $((\Qx.\psi,1t2,c),.,(\Qx.\psi,1t2),2,c)\in \delta_c$. Write the read value on to stack 2.
\item $((\Qx.\psi,1t2),.,(\Qx.\psi,2t1),1,x)\in \delta_c$. Write $x$ on stack 1 and change to copying back from Stack 2.
\item $((\Qx.\psi,2t1), c, (\Qx.\psi,2t1,c), 2)\in \delta_r$. Read and pop a value from stack 2.
\item $((\Qx.\psi,2t1,c),.,(\Qx.\psi,2t1),1,c)\in \delta_c$. Write the read value on to stack 1.
\item $((\Qx.\psi,2t1),\bot,\psi,2,\epsilon)\in \delta_c$. Copying is complete, move to the sub-formula.
\end{enumerate}

States where where the formula component  either begins with a $\forall x$ or 
has $\land$ as the outer most operator belongs to player $1$ and the other
states belongs to player $0$. 

In the second part we describe the state space starting at a state of the
form $(x = y)$ or $(x < y)$ that determines the winner of the game. This involves
additional states of the form $\{x = y, x < y, x, y a_y ~|~ x, y \in V \} \cup \{\True,\False\}$.  All these
positions belong to player $0$. 
The transitions (and states) are described as 
follows:

\begin{enumerate}
  \setlength{\itemsep}{1pt}
  \setlength{\parskip}{0pt}
  \setlength{\parsep}{0pt}
\item $(x = y, a, x=y,1) \in \delta_r$. Pop till  $x$ or $y$ are found.
\item $(x = y, z, x=y,1) \in \delta_r$, if $z \notin \{x,y\}$.
\item $(x = y, x, y,1) \in \delta_r$, start looking for $y$
\item $(x = y, y, x,1) \in \delta_r$, start looking for $x$
\item $(x, z, x,1) \in \delta_r$, skip other variables ($x \neq z$).
\item $(x, a, \False,1)$. Player 1 should win now.
\item $(x,x,\True,1)$. Player 0 should win now.
\item $(x < y, a, x<y, 1) \in \delta_r$. Pop till you find $x$.
\item $(x < y, z, x < y, 1) \in \delta_r$. Skip other variables ($z \notin \{x,y\}$).
\item $(x < y, y, \False,1) \in \delta_r$. Player $1$ wins.
\item $(x < y, x, a_y, 1)  \in \delta_r$. $x$ is seen first, make sure there is an $a$ before the $y$.
\item $(a_y, z, a_y, 1) \in \delta_r$, $z \neq y$.
\item $(a_y, a, \True, 1) \in \delta_r$. Player $0$ wins.
\item $(\True,\True) \in \delta_e$.
\item $(\False,\False) \in \delta_e$.
\end{enumerate}

It is quite easy to check that starting at a configuration of the form 
$(x = y, \gamma_1, \gamma_2)$, the play enters $\True$ iff the valuation
defined by $\gamma_1$ satisfies $x = y$ and similarly for $x > y$. Further
there is no phase change and every play eventually either enters $\True$
or $\False$.  The state $\True$ has parity $0$ ensuring victory for player
$0$ and state $\False$ has parity $1$.

However, starting at a configuration with quantifiers does not guarantee
that each play is terminating. This because of the \emph{loop} in states
of the form $(\Qx. \psi, >)$. However, we can make this unprofitable
for the owner by setting the parity to be a $0$ if  $\Qx = \forall x$ 
and setting the parity to be $1$ if $\Qx = \exists x$, thus forcing
the player to exit such states.  All other states are transient and hence their
parity does not matter and can be assigned anything. 

Thus, any winning strategy for either player in this game corresponds
to a winning strategy for the player in the satisfiability game.
Translating a winning strategy in the satisfiability game to a winning
strategy in this game is even easier. 
Further, observe that any run of this MPDS cannot
change phases more than 2 times the number of quantifiers in the formula and
thus it naturally defines a bounded phase game.   All this gives us the
following theorem.

\begin{theorem} \label{thm: reduction}
For any $\FO$ formula $\phi$ of size $n$  there is a MPDS game  with at most 
polynomial  states in $n$, for which the $2n$ bounded phase game  is equivalent to the
satisfiability game for $\phi$. Thus, solving parity games on bounded-phase MPDSs is non-elementary.
\end{theorem}
 
We also wish to remark that the alphabet of the MPDS need not grow with the
number of variables. We can encode the variables using two letters and this
will increase the state space (which will stay polynomial). Thus, the result
holds for fixed size alphabets as well.

\paragraph{Remark: }
In order to simply our presentation in the following sections, where the
constructions tend be much more involved, we shall often explain the role
of some subset of the state space in an informal manner when it is clear
how it can be formalized. For instance, instead of writing out the
state space beginning at $(x=y)$ above, we shall simply say that 
``there is a \emph{subroutine} beginning at a state $(x=y)$ that pops the
stack till it encounter $x$ or $y$ and then verifies that the other is
also encountered before any $a$'s and if so enters the state $\True$ and
otherwise the state $\False$. It is easy to see that the state space
needed for this subroutine is constant in size and it does not make
any phase (or context) changes''.

\section{MPDS, CTL and model checking}
In this section we show that model checking of bounded context-switch  MPDSs
w.r.t. CTL formulas has a non-elementary lower bound.
\subsection{The logic CTL}
The logic CTL is a simple temporal logic to describe branching time properties
of systems. The syntax of CTL is given by
$$ 
\alpha ~:=~ P ~\mid~ \alpha_1 \land \alpha_2 ~\mid~ \neg \alpha ~\mid~ \EX \alpha~ \mid~ \EF\alpha ~\mid~ \EG \alpha~\mid~ \alpha_1 \EU \alpha_2 
$$

where $P$ is a propositional variable drawn from a suitable set.

Models of CTL formulas are Kripke structures or LTSs. For our purposes
we may think of them as graphs where each node is labelled by the set
of propositions true at that node.  The formula $P$ is true at a state
$s$ if $P$ belongs to the label of $s$. The boolean operators have the
usual meaning. The formula $\EX \alpha$ is true at $s$ if there is an
edge to node $s'$ and $s'$ satisfies $\alpha$. $\EF \alpha$ is true at
$s$ if there is a reachable node $s'$ where $\alpha$ is true. $\EG \alpha$
asserts that there is a complete path (finite ending at a node with no
outgoing edges or infinite) such that every state appearing in that path
satisfies $\alpha$. Finally, $\alpha_1 \EU \alpha_2$ is satisfied at $s$
if there is a path $s=s_1,s_2, \ldots s_n$ such that $s_n$ satisfies
in $\alpha_2$ and $s_i$ satisfies $\alpha_1$ for $i < n$.

The model-checking problem for CTL is to determine for a given formula $\alpha$
and a labelled graph $G$ and a node $s$ whether $s$ satisfies $\alpha$.
For a formal semantics and  detailed introduction to CTL model-checking 
 may be found for instance in \cite{BaKa08,ClGrPe99}. 

We may turn any MPDS into a model by taking the set of control states as
the set of propositions with the obvious labeling -- $q$ is true only 
at the state $q$. The problem we consider is, given an MPDS $M$ and a CTL 
formula over its states $\alpha$, and a constant $k$,  restrict
its transition graph to at most $k$ context switches and  check if the 
initial configuration satisfies the formula $\alpha$. We call this
the \emph{bounded-context switch CTL model checking problem}.

Our main theorem is the following:

\begin{theorem}
Fix any constant $k$. The problem of model checking CTL 
formulas of size $m$ against MPDSs of size $n$ with a context bound $k$
has complexity that is at least $2^{2^{\ldots^{2^{P(m,n)}}}}$ where the
height of the tower is $g(k)$,  a linear function of $k$  and 
$P(m,n)$ is a polynomial in $m,n$.
\end{theorem}

\subsection{Stockmeyer's Nested Counters} 

Our proof draws heavily from the techniques developed by L. Stockmeyer in \cite{St74} and used heavily by Igor Walukiewicz and Thierry Cachat \cite{CaWa07} in showing that deciding reachability games for higher-order pushdown systems is non-elementary. We combine these with some ideas from a proof
of Igor Walukiewicz showing that model checking pushdown systems against
CTL formulas is \exptime-complete. In the rest of this section, we recall
some of these ideas from the aforementioned papers.

The number $\tower(k)$ is inductively defined as follows: $\tower(1) = 1$ and
$\tower(k) = 2^{\tower(k-1)}$ for $k > 1$. The function $\tower(k)$  grows
as a tower of exponents of $2$.
A key idea from \cite{CaWa07}  that we will need is that of a \emph{level $k$-counter}. 
These counters are parametrized by a natural number $n$. For instance when $n$ is $1$, 
a level $k$ counter stores a value in the range $0$ to $\tower(k) - 1$. In addition to storing a sequence of $\tower(k-1)$ bits  needed to
describe values in this range, a level $k$-counter also stores the address
of each of these bits using level $k-1$ counters.  

Let $\Sigma_i = \{a_i,b_i\}$, $i \geq 1$.   We also write $\Sigma^i$ for
$\bigcup_{j \leq i} \Sigma_j$.
The letters $a_i$ and $b_i$ are used to denote the
$0$ and $1$ values of the level $i$ counter respectively. We are now
in a position to formally define level $k$ counters.

\begin{definition}(\cite{CaWa07})
Fix an integer $n$.
\begin{itemize}
 \item A \emph{level $1$-counter}  is a word of length $n$ over the alphabet 
$\Sigma_1$. Thus interpreting $a_1$ and $b_1$ as $0$ and $1$ respectively,
the values that a $1$-counter takes varies from $0$ to $2^n-1$. The largest
value denoted by a level $1$ counter is denoted $\maxv{1}$ is $2^n - 1$.

\item A \emph{level  $k$-counter} is a word over the alphabet $\Sigma^k$ of the
form $l_0 \sigma_0,\cdots l_m \sigma_m$ with $\sigma_i \in \Sigma_k$ where,
each $l_i$ is a (k-1) level counter, $l_0$ is the  (k-1) level counter 
representation of the value $0$, $l_m$ represents  the value $\maxv{k-1}$.
and $\forall i < m, l_{i+1} = l_i +1$.
\end{itemize}
\end{definition}

We shall often write \emph{$k$ counter} to mean a level $k$ counter. Quite clearly,
$\maxv{k} = 2^{\maxv{k-1}}$.

\subsection{Coding Counters properties using MPDSs and CTL formulae}

Our lower bound construction involves maintaining configurations
of a bounded-space turing machine on the stacks of a multi-pushdown system.
The configurations are further encoded using the nested counters
 described in the previous section.  In order to achieve this
we need to be able to check certain basic properties regarding counters
and configurations stored on the stacks. In this section we address
the properties regarding counters and then follow it in the next section
with properties of configurations.
We intend to store the counters on the stack with the Most Significant Bit (MSB) on top of
stack.

\begin{definition}
\begin{enumerate}
\item $\first{k}$: \emph{Assuming that the top of the first counter contains a valid $k$ counter check that it has the minimum possible $k$ counter value.} 

Formally, a configuration $q(w_1,w_2)$ satisfies $\first{k}$ if $w_1 = l_i\sigma_1\gamma_1$, with $\sigma_1 \not \in \Sigma^k$ and $l_i$ is a valid $k$ counter
implies that every digit of $l_i$ is $a_k$ (denoting $0$).

\item $\last{k}$: \emph{Assuming that the top of the first counter contains a valid $k$ counter check that it has the maximum possible $k$ counter value.} 

Formally, a configuration $q(w_1,w_2)$ satisfies $\last{k}$ if $w_1 = l_i\sigma_1\gamma_1$, with $\sigma_1 \not \in \Sigma^k$ and $l_i$ is a valid $k$ counter
implies that every digit of $l_i$ is $b_k$ (denoting $1$).

\item $\Equal{k}$:  \emph{Assuming that the top of both the stacks contains  valid $k$ counters, check that these values are equal.} 

Formally, a configuration
$q(w_1,w_2)$ satisfies $\Equal{k}$ iff $w_1 = l_i\sigma_1\gamma_1$ and 
$w_2 = l_i\sigma_2\gamma_2$ with $l_i$ and $l'_i$  being a valid $k$ counters and $\sigma_1$
and $\sigma_2$ do not belong to $\Sigma^k$ implies that $l_i = l'_i$.

\item $\Succ{k}$: \emph{Assuming that the top of both the  stacks contains valid  $k$ counters, check that the value of on the second stack is the successor of the value
on the first stack.}

Formally, a configuration
$q(w_1,w_2)$ satisfies $\Succ{k}$ iff $w_1 = l_i\sigma_1\gamma_1$ and 
$w_2 = l'_i\sigma_2\gamma_2$ with $l_i, l'_i$  valid $k$ counters, 
$\sigma_1$ and $\sigma_2$ do not belong to $\Sigma^k$ implies
$l'_i$ is $l_i + 1$ and  

\item $\Valid{k}$:  \emph{Verify that the contents of the first stack begins
with a valid $k$ counter followed by some letter not in the alphabet $\Sigma^k$.}

Formally, a configuration
$q(w_1,w_2)$ satisfies $\Succ{k}$ iff $w_1 = l_i\sigma_1\gamma_1$, $l_i$ is
a valid $k$ counter and $\sigma_1 \not \in \Sigma^k$.
\end{enumerate}
\end{definition}

We shall next show that each of these properties can be ensured by the
addition of subroutines and restricting their behaviors via CTL formula
in a manner to be described below.

\subsubsection{Implementing $\last{k}$ and $\first{k}$}
We first add a new state, $\qmax{k}$ that pops the first stack 
till it encounters a letter outside $\Sigma^k$ and further enters the 
state $\error$ if it ever encounters the letter $a_k$ in doing so.  
Then, if there is an internal transition from a state $q$ to $\qmax{k}$ then,
a configuration $q(w_1,w_2)$, in which $w_1$ begins with a valid $k$
counter satisfies $\last{k}$ iff it does NOT 
satisfy the CTL formula $\ctlmax{k} = \EX (\qmax{k} \land \EF \error)$. 
One can implement $\first{k}$ quite similarly (using a state $\qmin{k}$
instead of $\qmax{k}$ and replacing $a_k$ by $b_k$.) 

Clearly this can be achieved by  an automaton with a constant number
of states (and $O(k)$ transitions since the alphabet depends on $k$) 
and it needs no context switches.  The size of the CTL formula is a 
constant. Across all the $k$ levels, we thus add $\Ord(k)$ states and 
make no context-switches.

\subsubsection{Implementing $\Equal{k}$.}
\paragraph{Simple Case:} $k = 1$\\
Remember that we need to check this only for configurations where both
stacks contain a valid $1$-counter,  i.e. a word of length $n$ over $\Sigma_1$,
on top. Add a subroutine, with new states,  that guesses 
a number $i \in \{0,\ldots,n-1\}$, pops $i$ symbols from both the stacks and
if the following symbols on the two stacks are different enters the state $\error$.

We can do this using at most one context switch. 
Pop the $i$ values from stack 1 before doing the same in stack $2$, 
maintaining a counter in the state that counts the number of pops 
on stack 1 so that we may pop the same number from the other stack.
The set of new states, denoted $\Qeqcheck{1}$ has  size $n \times 2$ 
(since we also need to remember the $i$th letter from stack
1 while popping stack 2). 
Let the starting state of this new subroutine be $\qeqcheck{1}$. 
Now, if there is an internal transition from a state $q$ to 
$\qeqcheck{1}$,  a configuration $q(w_1,w_2)$ is which $w_1$ and $w_2$
begin with valid $1$-counters satisfies $\Equal{1}$ iff 
it does NOT satisfy the CTL formula 
$\ctleq{1} = \EX ({\qeqcheck{1}} \land \EF \error)$. We also record the
fact that any run beginning at $\qeqcheck{1}$ makes at most one context switch.

Note that this subroutine has size  $\Ord(n)$ and makes at most $1$ context switch.
The size of the associated CTL formula is constant.

\paragraph{Induction:} 
The contents of the two stacks are of the form $l\sigma\gamma$ and
$l'\sigma'\gamma'$  and $l$ and $l'$ are valid $k$ counters. Thus
$l = l_{\maxv{k-1}}\sigma_{\maxv{k-1}} \ldots l_0\sigma_0$ and 
$l' = l'_{\maxv{k-1}}\sigma'_{\maxv{k-1}} \ldots l'_0\sigma'_0$. 
Since
the counters are well-formed it suffices to check that it is NOT the 
case that there is a $i$ and $j$ such that $l_i = l'_j$ and 
$\sigma_i \neq \sigma'_j$.   This ability to decouple the indices on
the two stacks is made possible by the special structure of the nested counters and permits us to bound the number of context switches needed.

Our subroutine begins, in a state $\qeqcheck{k}$,  by popping a number of words of the form  $c\sigma$, 
where $c \in (\Sigma^{k-1})^*$ and $\sigma \in \Sigma_k$  from stack 1.  This can be achieved by adding a constant number of states (but transitions linear in the alphabet and hence $k$.) It then removes a similar sequence 
(not necessarily of the same length) from stack 2, again requiring the 
addition of only constant number of states. Let the set of new states added be $\Qskip{k}$,
and we may assume w.l.o.g. that a successful run of this routine terminates in a 
state $\qskipfinal{k}$ which is entered for the first time at this point. 
Suppose, there is an internal transition from a state $q$ to the state $\qeqcheck{k}$, then 
starting at some configuration $q(l\sigma\gamma,l'\sigma'\gamma')$ 
a run of our subroutine will result in a configuration of the form
$\qskipfinal{k}(l_i\sigma_i l_{i-1}\sigma_{i-1} \ldots l_0\sigma_0\sigma\gamma,
l'_j\sigma'_j l'_{j-1}\sigma'_{j-1} \ldots l'_0\sigma'_0\sigma'\gamma')$.

We add internal transitions from $\qskipfinal{k}$  to $\qeqcheck{k-1}$ 
to verify whether $l_i = l'_i$.  We also add an internal transition from
$\qskipfinal{k}$ another state $\qrcchk{k-1}$.  

The subroutine, with state space $\Qrcchk{k-1}$ beginning at $\qrcchk{k-1}$ checks whether $\sigma_i = \sigma'_j$.  
It first pops a $k-1$ counter from stack 2 and then such a counter from
stack $1$ and enters $\error$ if the values following these in the two stacks
are different.  Again this can be done using at most 3 states 
and needs only one context switch.

If there is an internal transition from $q$ to $\qeqcheck{k}$ then 
a  configuration $q(l\sigma\gamma,l'\sigma'\gamma')$ with valid $k$ counters 
$l$ and $l'$ on top of the two stacks satisfies $\Equal{k}$ if and only
if it does NOT satisfy the CTL formula 
$\ctleq{k} ~=~  EX (\qeqcheck{k} \land EF (\qskipfinal{k} \land (\neg \ctleq{k-1} \land EX (\qrcchk{k-1} \land \EF \error))))$. 

The size of this subroutine, which includes the corresponding subroutine for
all values less than $k$,  is bounded by the sum of the size of the corresponding subroutine for $k-1$ (contributed by $\Qeqcheck{k-1}$)  and a constant dependent on $k$ (contributed by the states in $\Qskip{k} \cup  \Qrcchk{k-1}$).
Thus the size of $\Qeqcheck{k}$ is $\Ord(k^2 + n)$. 
Also observe that the maximum number of context switches possible is
$2$ plus the number of context-switches possible starting at $\qeqcheck{k-1}$.
Thus, the maximum number of context switches possible is $2*k$.
The size of the CTL formula $\ctleq{k}$  is $\Ord(k)$.

\subsubsection{Implementing $\Succ{k}$.}
We use once again use an observation used by Cachat-Walukiewicz. The binary
representation of the number $i+1$ can be obtained from that of $i$ as 
follows: Let $j$ be the first position, starting from the LSB, where a $0$
occurs in $i$.  Just flip all the bits in the positions up to $j$.
Thus, given the binary representations of two numbers $i$ and $\ell$, in order
to show that $i$ is not $\ell+1$, it suffices to either find a position between
the $j$ and  the LSB where the bits are identical or a position between the MSB
and $j+1$ that are different. We call such a position as a  \emph{faulty} position.

\paragraph{Base case:} $k=1$.
Pop  $j$ elements from stack $1$, $0 \leq j \leq n-1$. This is our guess of
the faulty position.   Remember $j$ in the state and pop $n-1-j$ more elements
 to learn whether to check for equality or inequality w.r.t. position $j$ in stack 2.  Then do the appropriate check on stack 2 entering the state $\error$ 
if $j$ is indeed a faulty position. The number of states added for this
subroutine is linear in $n$ and we use $\Qsuccheck{1}$ to denote this set 
and $\qsuccheck{1}$ to denote the initial state of this subroutine. Then, 
a configuration $q(l\sigma\gamma,l'\sigma'\gamma')$ satisfies $\Succ{1}$ 
iff it does NOT satisfy the CTL formula $\ctlsucc{1} = \EX (\qsuccheck{1} \land \EF\error)$.  

The number of states needed of $\Ord(n^2)$ and any run starting at the state $\qsuccheck{1}$ makes at most one context switch. Further, the size of the CTL
formula is constant.

\paragraph{Induction:} 
The contents of the two stacks are of the form $l\sigma\gamma$ and
$l'\sigma'\gamma'$  and $l$ and $l'$ are valid $k$ counters. Thus
$l = l_{\maxv{k-1}}\sigma_{\maxv{k-1}} \ldots l_0\sigma_0$ and 
$l' = l'_{\maxv{k-1}}\sigma'_{\maxv{k-1}} \ldots l'_0\sigma'_0$. 
Again, the structure of the construction remains the same. Repeat what was done for $k=1$ except that instead of counting out the 
position numbers in the two stacks use the addresses available in the 
nested counters.   

The subroutine begins by removing some sequence of address value pairs
from both the stacks (using at most one  context switch and needing only 3 states)
as in the case of equality check.  This phase ends in a state $\qskipfinalsuc{k}$.
At this point the configuration should be of the form 
$\qskipfinalsuc{k}(l_i\sigma_i l_{i-1}\sigma_{i-1} \ldots l_0\sigma_0\sigma\gamma,
l'_j\sigma'_j l'_{j-1}\sigma'_{j-1} \ldots l'_0\sigma'_0\sigma'\gamma')$.

There are internal transitions from $\qskipfinalsuc{k}$ to $\qeqcheck{k-1}$ to
check if $l_i = l'_j$ and  to three other states
\begin{itemize}
\item  $\qrcchk{k-1}$, which we have already seen in the previous subsection,
beginning a subroutine  which enters $\error$ only if $\sigma_i \neq \sigma'_j$.
\item $\qrcchkneq{k-1}$, beginning a subroutine which enters $\error$ only
if $\sigma_i = \sigma'_j$.
\item $\qscantype{k-1}$ which pops the remaining part of $l$ from the first 
stack entering $\qeq$ or $\qneq$ depending on whether there is a $m < i$
with $\sigma_m = a_k$ (i.e. $0$) or not.
\end{itemize}

Thus, $l$ is not $l'+1$ if and only if the subroutine $\qeqcheck{k-1}$ 
reports that the $l_i = l'_j$, and either $\qscantype{k-1}$ enters $\qeq$
and $\qrcchk{k-1}$ enters $\error$ or $\qscantype{k-1}$ enters $\qneq$ and
$\qrcchkneq{k-1}$ enters $\error$. 

Let $\Qsuccheck{k}$ be the set of new states added in the subroutine
described above. 
Let $\qsuccheck{k}$ 
be the initial state of this subroutine. 
Suppose there is an internal transition from a state $q$ to the 
state $\qsuccheck{k}$. Then, 
any configuration $q(l\sigma\gamma,l'\sigma'\gamma')$  satisfies $\Succ{k}$ 
if and only if it does NOT satisfy the CTL formula 
$\ctlsucc{k} = \EX (\qsuccheck{k} \land \EF (\qskipfinalsuc{k} \land (\neg \ctleq{k-1} \land (\EF\qeq \land \EX (\qrcchk{k-1} \land \EF \error)) \lor (\EF\qneq \land \EX (\qrcchkneq{k-1} \land \EF \error)))))$.

Observe that only a constant number of states are added
(the subroutine call to $\qeqcheck{k-1}$ does not create new states as
we may use the same copy used for the equality check). Thus, the size
of this subroutine is $\Ord(k + n^2)$. 
Once again we record that the number of context switches in any run
starting at $\qsuccheck{k}$ is bounded 2 plus the number of context 
switches from $\qeqcheck{k-1}$ and thus bounded $2*k$. Finally, observe
that the size of the CTL formula described above is $\Ord(k)$ since it
is a constant plus the size of the formula $\ctleq{k-1}$.

\subsubsection{Implementing $\Valid{k}$}

\paragraph{Base case:} $k = 1$.
It is sufficient to check that the stack contents begin with a sequence
of length $n$ over $\Sigma_1$  followed by a symbol not in $\Sigma_1$.
Our subroutine does this and enters the state $\error$ if this is not the case.
 Let $\Qvalid{1}$ be the
set of states and let $\qvalid{1}$ be the initial state of this subroutine.
If there is an internal transition from a state $q$ to the state $\qvalid{1}$
then, a configuration $q(w_1,w_2)$ satisfies $\Valid{1}$
iff it does NOT satisfy the CTL formula $\ctlval{1} = \EX (\qvalid{1} \land \EF \error)$.

We note that the size of $\Qvalid{1}$ is bounded by $n$ and routine performs
no context switches.  The size of the CTL formula is evidently constant.

\paragraph{Induction:} 
Suppose the configuration is $q(w,w')$. 
Let $w = l\sigma\gamma$ for some $l \in (\Sigma^k)^*$, $\sigma \not \in \Sigma^k$. 
Further let $l = l_m\sigma_ml_{m-1}\sigma_{m-1} \ldots l_0\sigma_0$ with 
$l_i \in (\Sigma^{k-1})^*$, $\sigma_i \in \Sigma_k$.
  We need to check that 
\begin{enumerate}
\item Each $l_j$ is a valid $k-1$ counter.
\item $l_m$ is the maximum possible $k-1$ counter (i.e. with a $b_{k-1}$ for
each digit.)
\item $l_0$ is the  minimum possible $k-1$ counter (i.e. with a $a_{k-1}$ for
each digit.)
\item For each $j>0$ $l_{j-1} + 1 = l_j$.
\end{enumerate}


In order to verify the first condition above, 
we set up a subroutine beginning at state $\qvalskip{k}$ which 
begins by popping a sequence belonging to $((\Sigma^{k-1})^* \Sigma_k)^*$ 
and then enters $\qvalinduct{k}$.  The state $\qvalinduct{k}$ 
has an internal transition to $\qvalid{k-1}$ Thus, if there is an internal
transition from $q$ to $\qvalid{k}$ then the configuration $q(w,w')$ satisfies
the first condition above iff it does NOT satisfy the CTL formula
$\ctlvalind{k} = \EX (\qvalskip{k} \land \EF (\qvalinduct{k} \land  \ctlval{k-1}))$. Also note that the subroutine beginning at $\qvalskip{k}$ adds only
a constant number of states and any run of this subroutine has at most
as many context switches as $\qvalid{k-1}$. The formula $\ctlvalind{k}$
has size bounded by a constant plus the size of $\ctlval{k-1}$.

Checking the second condition, assuming that the first condition is satisfied,
 corresponds to checking $\last{k-1}$. In effect, if $q$ has a internal
transition to $\qmax{k-1}$ and $q(w_1,w_2)$ is a configuration satisfying
property 1 then it satisfies property 2 iff it does NOT satisfy  the CTL formula
$\ctlvallast{k} = \ctlmax{k-1}$. This subroutine does not involve
any context-switches and adds only a constant number of states. The size
of the formula $\ctlvallast{k}$ is constant.

Again, assuming that the first condition is satisfied, checking the third
condition can be achieved using the subroutine that begins at state
$\qvalskipp{k}$ which begins by popping a sequence belonging to 
$((\Sigma^{k-1})^* \Sigma_k)^*$ and then enters $\qvalinductp{k}$, with
some $l_j$ on top of the stack.
The state $\qvalinductp{k}$ has internal transitions to $\qmin{k-1}$ (to
check $\first{k-1}$ holds for $l_j$ ) as well as to a state $\qchklast{k}$. The state
$\qchklast{k}$ pops a sequence of elements of $\Sigma^{k-1}$,
then pops an element of $\Sigma_k$ and verifies that the following letter
does not belong to $\Sigma^{k}$ and enters $\win$ on successfully carrying
out this task.  In effect the run from $\qchklast{k}$ ends at $\win$
iff $j = 0$. Thus, if a state $q$ has an internal transition to 
$\qvalskipp{k}$ and $q(w_1,w_2)$ is a configuration satisfying
the first two conditions then  it does NOT satisfy property 3  iff it satisfies
the CTL formula
$\ctlvalfirst{k} = \EX (\qvalskipp{k} \land \EF (\qvalinductp{k} \land (\neg \ctlmin{k-1}) \land \EX (\qchklast{k} \land \EF \win)))$. Once again, this
subroutine does not involve any context-switches and adds only a constant
number of states. The size of the formula $\ctlvalfirst{k}$ is constant.

Finally we describe how to check the fourth property assuming the first 
three are satisfied. Our strategy is the following. 
\begin{enumerate}
\item First pop a sequence belongining to $((\Sigma^{k-1})^*\Sigma_k)^*$ to 
guess a $j$ which violates property 4, that is $l_j \neq l_{j-1} + 1$. 
\item Copy $l_j$ to the other stack. 
\item Remove $l_j\sigma_j$ from the first stack.
\item Check for satisfaction of $\Succ{(k-1)}$.
\end{enumerate}
The tricky step is to copy $l_j$ on to second stack using few context-switches.
Once again we  use the power of combining subroutines with CTL assertions.
We set up a subroutine that writes down an arbitrary sequence over
$((\Sigma^{k-2})^*\Sigma_{k-1})^*$ in the second stack. We then 
  check (using the induction hypothesis) that it is a valid $k-1$ 
counter and that the resultant configuration satisfies $\Equal{(k-1)}$ to simulate
the effect of copying.

The subroutine begins at a state $\qvalskipr{k}$  which pops a sequence
from $((\Sigma^{k-1})^*\Sigma_k)^*$ and enters a state $\qguess{k-1}$.
When a run reaches this state the contents first stack would be $l_j\sigma_j l_{j-1}\sigma_{j-1} \ldots l_0\sigma_0\gamma$.
The subroutine beginning at $\qguess{k-1}$, empties the second stack if it
already is not empty and then  writes down an arbitrary sequence
over $((\Sigma^{k-2})^*\Sigma_{k-1})^*$ into the second stack 
 and enters a state $\qchkguess{k-1}$.   The state $\qchkguess{k-1}$ has
internal transitions to $\qvalid{k-1}(2)$\footnote{i.e. the start state of the 
subroutine that checks that at the top of stack 2, there is a valid $k-1$
counter, which can be constructed similar to our construction for stack 1} and to $\qeqcheck{(k-1)}$.  The state $\qchkguess{k-1}$ also
has a internal transition to the state $\qremovelj{k}$ which pops
the first stack till $l_j\sigma_j$ is removed and then enters a state 
$\qbeforesucccheck{(k-1)}$ which has an internal transition to  
$\qsuccheck{(k-1)}$.  

Assuming that the state $q$ has an internal transition to $\qvalskipr{k}$,
the configuration $q(w_1,w_2)$ satisfies the fourth property if and only
if it does NOT satisfy the following CTL property
$$
\begin{array}{lcl}
\ctlvalsucc{k} & = & \EX (\qvalskipr{k} \land \EF (\qguess{(k-1)} \land  \EF (  \qchkguess{(k-1)} \land (\neg \ctlval{(k-1)(2)}) \\
&& \land 
 (\neg \ctleq{(k-1)}) \land \EX (\qremovelj{k} \land \EF (\qbeforesucccheck{(k-1)} \land \ctlsucc{(k-1)}))))) \\
\end{array}
$$

The subroutine only contains a constant number of new states. The maximum number of context switches starting is $\qvalskipr{k}$ is bounded by 2 plus the maximum of the number of context switches made starting from $\qvalid{k-1}$, $\qsuccheck{(k-1)}$ and $\qeqcheck{(k-1)}$. Further, the size of the formula above
is constant plus the size of the formula $\ctlval{(k-1)}(2)$ and the size
of $\ctlsucc{(k-1)}$.

Finally we combine these four part into one. The state $\qvalid{k}$ has
internal transitions to $\qmax{k-1}$, $\qvalskip{k}$, $\qvalskipp{k}$ and
$\qvalskipr{k}$. Then, if $q$ is any state with an internal transition
to $\qvalid{k}$ then $q(w_1,w_2)$ satisfies the formula
$$\ctlval{k} ~=~  \EX (\qvalid{k} \land (\ctlvalind{k} \lor \ctlvallast{k} \lor  \ctlvalfirst{k} \lor  \ctlvalsucc{k})) $$ iff $w_1$ does not begin with a valid $k$ counter.

Summing the values from the four different cases, we note that the entire
subroutine only adds a constant number of new states. Thus, across all levels
$k$ the number of states added for this case is bounded by $\Ord(k + n)$. 
The maximum number
of context switches is bounded by the maximum of the number starting
at $\qvalid{k-1}$ and the number we get for case 4 above, which is indeed
higher. Thus the maximum number of context switches is bounded by $2*k$.
Finally, the size of the formula is $\Ord(2^k)$, since there are two
copies of $\ctlval{k-1}$ in the expression for $\ctlval{k}$ (one from the
first case and one from that last case).

Thus in total the subroutines built to handle the counter operations in 
this section need only $\Ord(k^2 + n^2)$ states. Further any call to any of these
subroutines makes at most $2*k$ context-switches and finally size of the
CTL formulas used in asserting the counter properties is bounded by 
$\Ord(2^k)$.

\subsection{ Turing Machines, MPDSs and CTL formulae}

We now show a method to encode configurations of a space bounded 
turing machine  with  an input of size $n$  and  at most $\maxv{k}$ tape 
cells using $k$ counters which are  stored and processed  using  the stacks of multi-pushdown
system.

Let $M = (Q_M, \Gamma_M, s_M,\delta_M,F_M)$ be such a turing machine.
The contents of the tape of such a machine may be written as a string of 
length $\maxv{k}$ over the alphabet $\Sigma_M = \Gamma_M \cup Q_M$, where a letter
from $Q_M$ occurs precisely once. We enrich this string by writing down
the address of each position of the string as a $k$ counter (Thus this
encoding looks like a $k+1$ counter except that the alphabet at level
$k+1$ is $\Sigma_M$ instead of $\{a_{k+1},b_{k+1}\}$. We call 
such a configuration a \emph{$k$ configuration} of $M$.

As in the case of $k$ counters we now show that it is possible check
certain properties regarding configurations that lie on top of the
stacks of a multipushdown system. 

\begin{definition}
\begin{enumerate}
\item \ValidConf{k} : The top of the first stack is of the form $\rho_1\zeta\gamma$ where $\rho_1$ is a valid $k$ configuration (with the right end of the tape on top) and $\zeta \not \in \Sigma_M \cup \Sigma^k$.

\item \InitConf{k}{w} : Assuming that the top of the first stack contains the encoding of some configuration followed by $\zeta$, verify that it is the initial
configuration on input $w$, where $w$ is of length $n$.

\item \FinalConf{k} : Assuming that the top of the first stack contains the encoding of some configuration followed by $\zeta$, verify that it is a final
configuration.

\item \EqConf{k} : Assuming that the top of the two stacks contain valid configurations $\rho_1$ and $\rho_2$ (followed by $\zeta$) verify that $\rho_1 = \rho_2$.

\item \SuccConf{k} : Assuming that Stack 1  begins with a valid $k$ configuration $\rho_1$ followed by $\zeta$ and that , stack 2  begins with a valid $k$ configuration $\rho_2$ followed by $\zeta$ verify that $\rho_1 \tmoves{M} \rho_2$.\footnote{To be precise, the configurations coded by $\rho_1$ and $\rho_2$ are related by $\tmoves{M}$ and not $\rho_1$ and $\rho_2$ themselves.}

\item \Move{k} : Assuming that the first stack contains two valid $k$ configurations one below the other and separated by a $\zeta$, (i.e.  it is of the 
form $\rho_1\zeta\rho_2\zeta\gamma$),  verify that $\rho_2 \tmoves{M} \rho_1$
\end{enumerate}
\end{definition}

We next show that each of these properties can be checked using special subroutines in combination with CTL formulae.

\subsubsection{Implementing \ValidConf{k}}
Suppose the configuration is $q(w,w')$.
Let $w = l\sigma\gamma$ for some $l \in (\Sigma^k \cup \Sigma_M)^*$, $\sigma \not \in \Sigma^k \cup \Sigma_M$.
Further let $l = l_m\sigma_ml_{m-1}\sigma_{m-1} \ldots l_0\sigma_0$ with
$l_i \in (\Sigma^{k})^*$, $\sigma_i \in \Sigma_M$.
  We need to check that
\begin{enumerate}
\item Each $l_j$ is a valid $k$ counter.
\item $l_m$ is the maximum possible $k$ counter (i.e. with a $b_{k}$ for
each digit.)
\item $l_0$ is the  minimum possible $k$ counter (i.e. with a $a_{k}$ for
each digit.)
\item For each $j>0$ $l_{j-1} + 1 = l_j$.
\item $\sigma = \zeta$.
\item Exactly one of the letter $\sigma_m, \sigma_{m-1} \ldots \sigma_0$ belongs  to $Q_M$.
\end{enumerate}

Observe that the first 4 properties are identical to those needed to check
the validity of counters  and we
omit the details. Items 5 and 6 constitute a simple regular property and 
we again omit the details.  Thus, we may construct a subroutine beginning
at a stat $\qvalidconf{k}$ that uses only constant number of new states
(and $\Ord(k + |\Sigma_M|)$ transitions) and which makes at most $2*k$ 
context switches on any run and a CTL formula $\ctlvalidconf{k}$, whose
size is $\Ord(2^k)$  such that, if $q$ is any state with an internal transition to $\qvalidconf{k}$ 
 then $q(w_1,w_2)$ does NOT satisfy the formula $\ctlvalidconf{k}$ if and only if $q(w_1,w_2)$
satisfies $\ValidConf{k}$.

\subsubsection{Implementing $\InitConf{k}{w}$, $\FinalConf{k}, \EqConf{k}$}

For configurations  $q(w_1,w_2)$ satisfying $\ValidConf{k}$,  the first
two properties are regular properties that can be checked easily  and hence
we omit the details.
Checking $\EqConf{k}$ can be done exactly as the equality of $k$ counters was checked and the details are omitted.

We assume the presence of subroutines beginning at $\qinit{k}{w}$, $\qfinal{k}$ and $\qeqconfig{k}$, CTL formulas $\ctlinitconf{k}$, $\ctlfinalconf{k}$ and 
$\ctleqconf{k}$ such that if $q$ is any state with an internal transition
to $\qinit{k}{w}$ or $\qfinal{k}$ or $\qeqconfig{k}$ then it does NOT satisfy
$\ctlinitconf{k}$ or $\ctlfinalconf{k}$ or $\ctleqconf{k}$ 
iff it satisfies $\InitConf{k}{w}$ or $\FinalConf{k}$ or $\EqConf{k}$ respectively.

In the case of $\InitConf{k}{w}$ the number of states added is $\Ord(|w|)$ 
and in all the other cases we only add a constant number of new states, and hence $\Ord(k)$ across all the levels and  any of these subroutines
makes at most $2*k$ context switches and the sizes of the formula are in
$\Ord(k+|\Sigma_M|)$.

\subsubsection{Implementing $\SuccConf{k}$}
We assume that the TM in each move either modifies the current tape cell
or moves (left or right). So, if $C_1 = x_1 a q b x_2$ is a configuration
and $C_1 \tmoves{M} C_2$ then $C_2 = x_1 d e f x_2$. A move changes at most
2 positions, the position where $Q_M$ appears and one of its adjacent
positions. Thus to check if $C_2$ is reachable from $C_1$ by a move
it suffices to check that firstly, all positions that are at distance
$2$ or more from an element of $Q_M$ are unchanged, and the segment
of length three with an element of $Q_M$ in the middle is transformed
in accordance with a move. 

Let $\rho_1 = l_m \sigma_m l_{m-1} \sigma_{m-1} \ldots l_0 \sigma_0$ and
$\rho_2 = l_m \sigma'_m l_{m-1} \sigma'_{m-1} \ldots l_0 \sigma'_0$.  
This construction is similar to the construction for checking $\Succ{k}$ 
and we set up subroutines that try to check if one of the two properties
mentioned above is violated.

The state $\qsimple{k}$ begins a subroutine that first removes an
element of $((\Sigma^k)^* \Sigma_M)^*$ from stack 1 ensuring that the last element
removed is not an element of $Q_M$. It then enters a state $\qsimplef{k}$ 
which has an internal transition to  states $\qnotfromq{k}$  and 
$\rsimple{k}$. 

$\rsimple{k}$ removes an element of $((\Sigma^k)^* \Sigma_M)^*$ from stack 2
and enters a state $\rsimplef{k}$. Starting with $\qsimple{k}(\rho_1\zeta\gamma_1,\rho_2\zeta\gamma_2)$ in the stack, a run that reaches $\rsimplef{k}$  will
result in a configuration of the form 
$\rsimplef{k}(l_i \sigma_i l_{i-1}\sigma_{i-1} \ldots l_0 \sigma_0,l_j\sigma'_j l_{j-1}\sigma'_{j-1} \ldots l_0\sigma'_0)$  and by construction $\sigma_{i+1} \in \Gamma_M$.  
$\rsimplef{k}$  has internal transitions to the states $\qeqcheck{k}$ and
to the state $\qrcchkn{k}$. The subroutine is at $\qrcchkn{k}$ removes
a $k$-counter from both the stacks and enters the state $\error$ if the following
symbol on stack 1 is not in $Q_M$ and different from the next symbol on stack 2.
 Thus, in the configuration referred to above,
the formula $\neg\ctleq{k} \land \EX (\qrcchkn{k} \land \EF \error)$ witnesses
the fact that $i = j$ and $\sigma_i \neq \sigma'_j$.

$\qnotfromq{k}$ removes a $k$ counter and enters $\win$ if the next value
is not an element of $Q_M$ and its role is to verify that $\sigma_{i-1}$ is
not an element of $Q_M$. Thus the configuration $\qsimple{k}(\rho_1\zeta\gamma_1,\rho_2\zeta\gamma_2)$ satisfies the CTL formula 
$$ \ctlsimple{k} = \EF (\qsimplef{k} \land \EX (\qnotfromq{k} \land \EF \win) \land \EX (\rsimple{k} \land \EF (\rsimplef{k} \land \neg\ctleq{k} \land \EX(\qrcchkn{k} \land \EF \error))))$$
only if there is a position $i$ such that $\sigma_{i-1},\sigma_i,\sigma_{i+1} \not \in Q_M$ and $\sigma_i \neq \sigma'_i$.

The subroutine starting at $\qsimple{k}$ adds only a constant number of
new states, and the maximum number of context-switches is along the path via $\rsimple{k}$ leading to  $\qeqcheck{k}$ and is thus bounded by $2 + 2*k$.
The size of the formula $\ctlsimple{k}$ is at most $\Ord(k)$. 

To handle the three positions at distance $\leq 1$ from the position with
an element of $Q_M$ we have a subroutine beginning at state $\qhard{k}$. 
The state $\qhard{k}$ pops a sequence from $((\Sigma^k)^* \Sigma_M)^*$ from stack 1  and
enters a state $\rhard{k}$.  $\rhard{k}$ removes a sequence from
$((\Sigma^k)^*\Sigma_M)^*$ from stack 2   and enters the state $\rhardf{k}$.
Starting with $\qhard{k}(\rho_1\zeta\gamma_1,\rho_2\zeta\gamma_2)$ in the stack, a run that reaches $\rhardf{k}$  will
result in a configuration of the form 
$\rhardf{k}(l_i \sigma_i l_{i-1}\sigma_{i-1} \ldots l_0 \sigma_0,l_j\sigma'_j l_{j-1}\sigma'_{j-1} \ldots l_0\sigma'_0)$. 
$\rhardf{k}$ has internal transitions to the state $\qeqcheck{k}$, $\qtrymoves{k}$  and the state $\rtrymoves{k}$.

The role of $\qtrymoves{k}$ and $\rtrymoves{k}$ is to identify the letters
at the 3 positions at distance $\leq 1$ from the state. $\qtrymoves{k}$ 
has internal transitions to states $q_{(a,q,b)}$ where $a,b \in \Gamma_M$
and $q \in Q_M$. $q_{(a,q,b)}$ pops the elements of stack 1 and enters the
state $\win$ iff the first three elements of $\Sigma_M$ it removes are
$a$, $q$ and $b$ respectively. The behavior of $\rtrymoves{k}$ and $r_{(a,b,c)}$ is similar (where $a,b,c \in \Sigma_M$ and exactly one of them belongs
to $Q_M$. 

Let $V = \{((a,q,b),(d,e,f) ~|~ aqb \not \tmoves{M} def\}$.  The configuration
$\qhard{k}(\rho_1\zeta\gamma_1, \rho_2\zeta\gamma_2)$ satisfies the CTL formula 
$$ \ctlhard{k} = \EF (\rhardf{k} \land \neg \ctleq{k} \land \bigvee_{((a,q,b),(d,e,f)) \in V} (\EX (q_{(a,q,b)} \land \EF \win) \land \EX (r_{(d,e,f)} \land \EF \win))) $$
iff the three positions in $\rho_1$ around the occurrence of the state do 
not entail the corresponding positions in $\rho_2$ through any valid move.

The subroutine starting at $\qhard{k}$ adds $\Ord(|\Sigma_M|^3)$ 
states, and the maximum number of context-switches is along the path via $\rhard{k}$ leading to to $\qeqcheck{k}$ and is thus bounded by $2 + 2*k$.
The size of the formula $\ctlhard{k}$ is at most $\Ord(k + |\Sigma_M|^6)$.

Let $\qmove{k}$ be a state with internal transitions to $\qsimple{k}$ and
$\qhard{k}$.  If a state $q$ has an internal transition to $\qmove{k}$ then
the configuration $q(\rho_1\zeta\gamma_1,\rho_2,\zeta,\gamma_2)$ satisfies
the CTL formula 
$$ \ctlmove{k} = \EX (\qmove{k} \land (\EX(\qsimple{k} \land \ctlsimple{k})  \lor \EX(\qhard{k} \land \ctlhard{k})))$$ 
iff $\rho_1 \not \tmoves{M} \rho_2$.

The total number of states added therefore is bounded by $\Ord(|\Sigma_M|^3)$, the number of context-switches bounded by $2 + 2*k$ and the size
of the formula is bounded by $\Ord(k + |\Sigma_M|^6)$.

\subsubsection{Implementing \Move{k}}

Having implemented $\EqConf{k}$ and $\SuccConf{k}$, implementing $\Move{k}$ is not difficult. The idea is to \emph{copy} the first configuration on to the 
second stack (using a similar idea to the one used in $\Valid{k}$)  by 
generating an arbitrary sequence, and testing that it is valid (using $\ValidConf{k}$) and correct (using  $\EqConf{k}$). 
Then, we remove one configuration from stack 1 and then  we use 
use $\SuccConf{k}$ to verify whether the copy on the second stack is indeed
the reachable by a move from the configuration on top of the first stack. 
The details are as follows.

The subroutine beginning at the state $\qstep{k}$ empties the second
stack and writes down an arbitrary sequence from $(\Sigma^k \cup \Sigma_M)^* \zeta$ and enters $\qguessconfchk{k}$. The state $\qguessconfchk{k}$ has internal
transitions to the states $\qvalidconf{k}(2)$\footnote{Once again, a variant that checks that the value in Stack 2, instead of Stack 1, is a valid $k$ configuration} and $\qeqconfig{k}$. $\qguessconfchk{k}$ also has an internal transition
to $\qrcmove{k}$. The subroutine beginning at $\qrcmove{k}$ removes the top
of the first stack up to (and including) the first $\zeta$ and enters the state
$\qrcmovef{k}$ which in turn has an internal transition to $\qmove{k}$.  

Then, any state $q$ with an internal transition to $\qmove{k}$, a configuration
$q(\rho_1\zeta\rho_2\zeta\gamma_1,\gamma_2)$  satisfies the
CTL formula 
$$\ctlstep{k} =  \EX (\qstep{k} \land \EF (\qguessconfchk{k} \land \neg \ctlvalidconf{k}(2) \land \neg \ctleqconf{k} \land \EX (\qrcmove{k} \land \EF (\qrcmovef{k} \land \ctlmove{k})))) $$

iff it does not satisfy $\Move{k}$.

We add only a constant number of new states here. The maximum number of
context switches is bounded by the maximum of $1 + 2*k$ (for the path through
$\qvalidconf{k}(2)$), $1 + 2*k$ (for the path through $\qeqconfig{k}$) and
$2 + 2 + 2*k$ for the path through $\qmove{k}$. Thus the maximum number
of context switches is bounded by $4 + 2*k$. The size of the formula
$\ctlstep{k}$ is $\Ord(2^k + |\Sigma_M|^6)$.

Thus overall, across the subroutines for the counters and configurations
we have added only a $\Ord(n^2 + k^2 + |\Sigma_M|^3)$ states, make
at most $4 + 2*k$ context-switches in any run and any formula used is
bounded in size by $\Ord(2^k + |\Sigma_M|^6)$.

\subsection{From Space Bounded TMs to Model-Checking MPDSs}

In this section we utilize the constructions of the previous two sections
to show that for any given TM  $M$ working nondeterministic space $\tower(k)$ and
a input word $w$ of length $n$, we can construct a MPDS $A$ whose state space 
is polynomial in $n$, $k$ and the size of $M$,  a CTL formula
$\alpha$, both whose size is polynomial in the size of $M$, $w$ and exponential
in the size of $k$,  such that the MPDS $A$ 
makes at most $2*k + 5$ context switches in any run and 
 $A$ satisfies the formula $\alpha$ iff the TM has
an accepting run on the word $w$. 
 Thus, model-checking of MPDSs under the bounded context-switch  restriction against CTL formulas has a non-elementary lower-bound.

The idea is quite simple. The MPDS writes down a sequence of $\zeta$ separated
strings that could each potentially be a $k$ configuration. We 
use the techniques of the previous section to verify that each such 
string is a valid $k$ configuration and that it can be reached by a move
from the previously written configuration.  We also check that
the first configuration it writes down is the initial configuration on $w$  
and that it eventually writes a final configuration. Clearly,  all of this is 
possible only if the given Turing machine has an accepting run on $w$.

The MPDS we construct works as follows. It starts a state $\qstart{k}$ 
with just the $\bot$ in both stacks. The state $\qstart{k}$ begins
a subroutine which writes down a sequence in 
$\zeta(\Sigma^k \cup \Sigma_M)^*$ and then enters a
state $\qinitcheck{k}{w}$. This state $\qinitcheck{k}{w}$ has internal transitions
to the states $\qvalidconf{k}$ and $\qinit{k}{w}$.  
The state $\qinitcheck{k}$ also has an internal 
transition to a state $\qwrite{k}$. The state  $\qwrite{k}$ begins a routine 
which writes down a sequence in $\zeta(\Sigma^k \cup \Sigma_M)^*$ and enters
the state $\qmovecheck{k}$. The state $\qmovecheck{k}$ has an internal 
transition to $\qvalidconf{k}$, $\qstep{k}$, $\qfinal{k}$ and to $\qwrite{k}$ as well.

This system satisfies the CTL formula
$$ \begin{array}{l}\ctlaccept{k}{w} = \qstart{k} \land \EF ((\qinitcheck{k}{w}  \land \neg \ctlinitconf{k}{w}) \land \\
~~~~~~~((\qmovecheck{k} \Rightarrow (\neg \ctlvalidconf{k} \land \neg \ctlstep{k})) ~\EU~ (\qmovecheck{k} \land \neg \ctlvalidconf{k} \land \neg \ctlstep{k} \land \neg \ctlfinalconf{k})))
\end{array}$$
iff the turing machine has an accepting run.

The number of states added is constant (and $\Ord(k + |\Sigma_M|)$ transitions are added). The maximum number of context-switches is through $\qmovecheck{k}$ and then via $\qstep{k}$  and is bounded by
$1 + 4 + 2*k$. The size of the CTL formula above is bounded by 
$\Ord(2^k + |\Sigma_M|^6)$.

\subsubsection{Eliminating $\EU$}

We now show that  actually we can restrict ourselves to the fragment of CTL
consisting of $\EX$ and $\EF$ and still obtain the same lowerbound.
For this we modify the construction described above slightly. The automaton
first writes down an entire  sequence of potential configurations and 
then  checks that it is a valid accepting run, instead of doing so as 
each configuration is generated. The details are as follows.

 Now, the
MPDS writes down a sequence of words from $\zeta(\Sigma^k \cup \Sigma_M)^*$  
on the stack (starting at state $\qstart{k})$ and then enters a state
$\qruncheck{k}$. The state $\qruncheck{k}$ has an internal transitions to 
$\qvalidconf{k}$, $\qfinal{k}$, $\qstep{k}$  and $\qremoveconf{k}$.
The state $\qremoveconf{k}$ repeatedly removes ane element of 
$(\Sigma^k \cup \Sigma_M)^*\zeta$ and re-enters itself. 
 The state $\qremoveconf{k}$  also has internal transitions to $\qvalidconf{k}$,
$\qstep{k}$,$\qinit{k}{w}$,  $\qonemore{k}$ and  $\qtwomore{k}$.

The state $\qonemore{k}$ attempt to remove a sequence form $(\Sigma^k \cup \Sigma_M))^+\zeta$ and enters the state $\win$ if it succeeds. The state $\qonemore{k}$ does the same if it succeeds in removing two such sequences.
Then the MPDS satisfies the following CTL formula $\ctlacceptp{k}{w}$ iff  
the TM accepts the word $w$.  
$$ 
\begin{array}{l}
\ctlacceptp{k}{w} = \qstart{k} \land \EF ( \\
~~~~(\qruncheck{k} \land \neg \ctlvalidconf \land \neg  \ctlstep{k} \land \neg \ctlfinalconf{k}) \\
~~~~  \land \neg \EF (\qremoveconf{k} \land \EX (\qonemore{k} \land \EF\win) \land \ctlvalidconf{k}) \\
~~~~ \land \neg \EF (\qremoveconf{k} \land \EX (\qtwomore{k} \land \EF\win) \land \ctlstep{k})\\
~~~~ \land \neg \EF (\qremoveconf{k} \land \EX (\qonemore{k} \land \EF\win) \land
\EX (\qtwomore{k} \land \neg \EF \win) \land \ctlinitconf{k}{w})\\
) 
\end{array}
$$

This construction adds only a constant number of new states (and $\Ord(k + |\Sigma_M|)$ transitions), makes at most
$4 + 2*k$ context switches and the size of the formula $\ctlacceptp{k}{w}$ is
$\Ord(|w| + |\Sigma_M| + 2^k)$.

In summary, given a Turing machine $M$ and a word $w$ we can construct
a MPDS $A$ with state space $\Ord(|w|+|\Sigma_M|^3 + k)$ which makes at most $4 + 2*k$
context switches and a formula $\alpha$, whose size is $\Ord(|w| + |\Sigma_M|^6 + 2^k)$, such that $A$ satisfies $\alpha$ iff $M$ accepts $w$ in space $2^{2^{{\ldots}^{|w|}}}$ where the height of the tower is $k$. 


\paragraph{Observation :}
It is also possible reduce Alternating
Turing Machines instead of Nondeterministic machines, but the additional
work does not buy us much. 
$$ ASPACE (\tower(k/2))  \subseteq DTIME(\tower(k/2+1))  \subseteq  DSPACE(\tower(k/2+1))  \subseteq NSPACE(\tower(k/2+1)) $$ 
So, we just get to  increase the height of the tower by 1.

\bibliographystyle{abbrv}
\bibliography{/Users/kumar/tiger/Papers/Bibiliography}

\begin{thebibliography}{10}

\bibitem{At10a}
M.~F. Atig.
\newblock From multi to single stack automata.
\newblock In P.~Gastin and F.~Laroussinie, editors, {\em CONCUR}, volume 6269
  of {\em Lecture Notes in Computer Science}, pages 117--131. Springer, 2010.

\bibitem{At10b}
M.~F. Atig.
\newblock Global model checking of ordered multi-pushdown systems.
\newblock In Lodaya and Mahajan \cite{DBLP:conf/fsttcs/2010}, pages 216--227.

\bibitem{AtBoHa08}
M.~F. Atig, B.~Bollig, and P.~Habermehl.
\newblock Emptiness of multi-pushdown automata is 2etime-complete.
\newblock In M.~Ito and M.~Toyama, editors, {\em Developments in Language
  Theory}, volume 5257 of {\em Lecture Notes in Computer Science}, pages
  121--133. Springer, 2008.

\bibitem{AtBoQa09}
M.~F. Atig, A.~Bouajjani, and S.~Qadeer.
\newblock Context-bounded analysis for concurrent programs with dynamic
  creation of threads.
\newblock In S.~Kowalewski and A.~Philippou, editors, {\em TACAS}, volume 5505
  of {\em Lecture Notes in Computer Science}, pages 107--123. Springer, 2009.

\bibitem{AtTo09}
M.~F. Atig and T.~Touili.
\newblock Verifying parallel programs with dynamic communication structures.
\newblock In S.~Maneth, editor, {\em CIAA}, volume 5642 of {\em Lecture Notes
  in Computer Science}, pages 145--154. Springer, 2009.

\bibitem{BaKa08}
C.~Baier and J.-P. Katoen.
\newblock {\em Principles of Model Checking (Representation and Mind Series)}.
\newblock The MIT Press, 2008.

\bibitem{BrChCiCr96}
L.~Breveglieri, A.~Cherubini, C.~Citrini, and S.~Crespi-Reghizzi.
\newblock Multi-push-down languages and grammars.
\newblock {\em Int. J. Found. Comput. Sci.}, 7(3):253--292, 1996.

\bibitem{Ca03a}
T.~Cachat.
\newblock Higher order pushdown automata, the caucal hierarchy of graphs and
  parity games.
\newblock In J.~C.~M. Baeten, J.~K. Lenstra, J.~Parrow, and G.~J. Woeginger,
  editors, {\em ICALP}, volume 2719 of {\em Lecture Notes in Computer Science},
  pages 556--569. Springer, 2003.

\bibitem{CaWa07}
T.~Cachat and I.~Walukiewicz.
\newblock The complexity of games on higher order pushdown automata.
\newblock {\em CoRR}, abs/0705.0262, 2007.

\bibitem{CaWo03}
A.~Carayol and S.~W{\"o}hrle.
\newblock The caucal hierarchy of infinite graphs in terms of logic and
  higher-order pushdown automata.
\newblock In {\em FSTTCS'03}, pages 112--123, 2003.

\bibitem{ClGrPe99}
E.~M. Clarke, Jr., O.~Grumberg, and D.~A. Peled.
\newblock {\em Model checking}.
\newblock MIT Press, Cambridge, MA, USA, 1999.

\bibitem{HaTo10}
M.~Hague and A.~W. To.
\newblock The complexity of model checking (collapsible) higher-order pushdown
  systems.
\newblock In Lodaya and Mahajan \cite{DBLP:conf/fsttcs/2010}, pages 228--239.

\bibitem{HeLeMuSu10}
A.~Heu{\ss}ner, J.~Leroux, A.~Muscholl, and G.~Sutre.
\newblock Reachability analysis of communicating pushdown systems.
\newblock In C.-H.~L. Ong, editor, {\em FOSSACS}, volume 6014 of {\em Lecture
  Notes in Computer Science}, pages 267--281. Springer, 2010.

\bibitem{Ka09}
V.~Kahlon.
\newblock Boundedness vs. unboundedness of lock chains: Characterizing
  decidability of pairwise cfl-reachability for threads communicating via
  locks.
\newblock In {\em LICS}, pages 27--36. IEEE Computer Society, 2009.

\bibitem{LaRe09}
A.~Lal and T.~W. Reps.
\newblock Reducing concurrent analysis under a context bound to sequential
  analysis.
\newblock {\em Formal Methods in System Design}, 35(1):73--97, 2009.

\bibitem{LaToKiRe08}
A.~Lal, T.~Touili, N.~Kidd, and T.~W. Reps.
\newblock Interprocedural analysis of concurrent programs under a context
  bound.
\newblock In Ramakrishnan and Rehof \cite{DBLP:conf/tacas/2008}, pages
  282--298.

\bibitem{DBLP:conf/fsttcs/2010}
K.~Lodaya and M.~Mahajan, editors.
\newblock {\em IARCS Annual Conference on Foundations of Software Technology
  and Theoretical Computer Science, FSTTCS 2010, December 15-18, 2010, Chennai,
  India}, volume~8 of {\em LIPIcs}. Schloss Dagstuhl - Leibniz-Zentrum fuer
  Informatik, 2010.

\bibitem{MaPa11}
P.~Madhusudan and G.~Parlato.
\newblock The tree width of auxiliary storage.
\newblock In T.~Ball and M.~Sagiv, editors, {\em POPL}, pages 283--294. ACM,
  2011.

\bibitem{QaRe05}
S.~Qadeer and J.~Rehof.
\newblock Context-bounded model checking of concurrent software.
\newblock In N.~Halbwachs and L.~D. Zuck, editors, {\em TACAS}, volume 3440 of
  {\em Lecture Notes in Computer Science}, pages 93--107. Springer, 2005.

\bibitem{DBLP:conf/tacas/2008}
C.~R. Ramakrishnan and J.~Rehof, editors.
\newblock {\em Tools and Algorithms for the Construction and Analysis of
  Systems, 14th International Conference, TACAS 2008, Held as Part of the Joint
  European Conferences on Theory and Practice of Software, ETAPS 2008,
  Budapest, Hungary, March 29-April 6, 2008. Proceedings}, volume 4963 of {\em
  Lecture Notes in Computer Science}. Springer, 2008.

\bibitem{Se09}
A.~Seth.
\newblock Games on multi-stack pushdown systems.
\newblock In S.~N. Art{\"e}mov and A.~Nerode, editors, {\em LFCS}, volume 5407
  of {\em Lecture Notes in Computer Science}, pages 395--408. Springer, 2009.

\bibitem{Se10}
A.~Seth.
\newblock Global reachability in bounded phase multi-stack pushdown systems.
\newblock In T.~Touili, B.~Cook, and P.~Jackson, editors, {\em CAV}, volume
  6174 of {\em Lecture Notes in Computer Science}, pages 615--628. Springer,
  2010.

\bibitem{St74}
L.~J. Stockmeyer.
\newblock {\em The complexity of decision problems in automata theory and
  logic.}
\newblock PhD thesis, M.I.T., Cambridge, Massachusetts, U.S.A., 1974.

\bibitem{ToMaPa07}
S.~L. Torre, P.~Madhusudan, and G.~Parlato.
\newblock A robust class of context-sensitive languages.
\newblock In {\em LICS}, pages 161--170. IEEE Computer Society, 2007.

\bibitem{ToMaPa08a}
S.~L. Torre, P.~Madhusudan, and G.~Parlato.
\newblock Context-bounded analysis of concurrent queue systems.
\newblock In Ramakrishnan and Rehof \cite{DBLP:conf/tacas/2008}, pages
  299--314.

\bibitem{ToNa11}
S.~L. Torre and M.~Napoli.
\newblock Reachability of multistack pushdown systems with scope-bounded
  matching relations.
\newblock In J.-P. Katoen and B.~K{\"o}nig, editors, {\em CONCUR}, volume 6901
  of {\em Lecture Notes in Computer Science}, pages 203--218. Springer, 2011.

\bibitem{Wa00}
I.~Walukiewicz.
\newblock Model checking ctl properties of pushdown systems.
\newblock In S.~Kapoor and S.~Prasad, editors, {\em FSTTCS}, volume 1974 of
  {\em Lecture Notes in Computer Science}, pages 127--138. Springer, 2000.

\bibitem{Wa01}
I.~Walukiewicz.
\newblock Pushdown processes: Games and model-checking.
\newblock {\em Inf. Comput.}, 164(2):234--263, 2001.

\bibitem{Wo05}
S.~Wohrle.
\newblock {\em Decision Problems over Infinite Graphs: Higher-order Pushdown
  Systems and Synchronized Products}.
\newblock PhD thesis, RWTH-Aachen University, 2005.

\end{thebibliography}

\end{document}